# Elaboration and characterization of bioplastic films based on bitter cassava starch (Manihot esculenta) reinforced by chitosan extracted from crab (Shylla seratta) shells

[1]Research Scholar Julie Tantely Mitantsoa, [2]Pr. Pierre Hervé Ravelonandro, [3]Research Scholar. Fara Arimalala And rianony, [4]Dr. Rajaona Rafihavanana And rianaivoravelona

[1,2,3,4]Department of Processes and Industrial Ecology, Laboratory Unité de Recherche en Génie de l'Eau et Génie de l'Environnement (URGPGE), Faculty of Sciences, University of Antananarivo, Madagascar
Corresponding author's-mail: mitantsoajulietantely@gmail.com

**Abstract- Bioplastics are polymer plastics which are derived from renewable biomass resources. In this study, bioplastic films based on two different polysaccharides such as bitter cassava starch and chitosan extracted from crab shells were produced by casting technique, using glycerol as plasticizer. The purposes of this research are to characterize and to figure out the effect of additional chitosan concentrations (0; 10; 20; 30; 50% by weight of starch) on the physicochemical, mechanical and water barrier properties of bioplastic films. The film's solubility in water (S), water absorption capacity (WAP), water vapor permeability (WVP), tensile strength (TS), elongation at break (E), Young's modulus (YM) and biodegradability were investigated. The possible interactions between starch and chitosan molecules were evaluated by Fourier transform infrared spectroscopy (FTIR). From the analysis, the incorporation of the chitosan shows improved results on the water barrier properties of the bioplastic films. Optimum solubility in water, water absorption capacity, and water vapor permeability are obtained on the composition of starch/chitosan was 50/50. Actually, the addition of chitosan increased tensile strength, and elongation at break. The characterization of optimum mechanical proprieties also occurred on the 50/50 composition of cassava-starch and chitosan. At this ratio, tensile strength obtained were 6,3000 MPa; and the elongation at break were 62,8571%. It was found that cassava-starch/chitosan-based films have a stable structure compared to native cassava-starch films.**

**Keywords- Chitosan, Thermoplastic starch, Starch-chitosan films, Bioplastic, Mechanical properties, and Biodegradable films**

## I.INTRODUCTION

Recently, environmental destruction has been one of the major issues around the world. This is mainly manifested by the degradation of the ozone layer, acid rain, global warming, increasing

Green house gases, air pollution, water pollution and soil pollution. Among the causes which accelerate this destruction of our environment, we cannot neglect the impacts caused by the use of traditional plastics from petrochemicals raw materials.

Once unusable, they end up as waste and have negative effects on the environment. They take several years to completely decompose because they cannot be degraded under the influence of solar radiation or microbial decomposers [1] [2]. Their biological, physical and chemical resistance causes significant pollution of nature on land and especially in the oceans.

Thus, the use of natural biopolymers that are easily biodegradable and renewable would solve these problems. It is in this context that we have sought a renewed interest in the development of biodegradable materials from biomass in order to reduce environmental pollution due to the use of non-biodegradable plastic bags by proceeding with the search for alternatives.

Amidst natural biopolymers, starch is considered as the most promising raw material for the development of new materials that are more environmentally friendly. However, there are limitations for the use of starchy materials compared to petrochemical materials due to the high moisture absorption caused by their hydrophilic character, which provokes poor mechanical and barrier properties leading to low stability [3]. Mixing cassava-starch films with other bio-based polymers such as chitosan has been reported to be a good technique for achieving fully bio-based materials [4]. Thus, improving these physical and mechanical properties in order to widen the application window of thermoplastic starch [4] [5].

Chitosan is a natural biopolymer produced from chitin which has been found in a wide range of natural sources such as crustaceans. It is a linear polysaccharide composed essentially of β-(1,4)-linked N-acetyl-D-glucosamine [6] [5]. Chitin is made up of linearchain of acetylglucosamine groups while chitosan is obtained by removing enough acetyl groups for the molecule to be soluble in most diluted acids with pH less than 6 due to the protonation of its amino groups. Chitosan is actually characterized by its biodegradability, biocompatibility and non-toxicity [7].

Indeed, this study aims to develop and characterize bioplastic films including a mixture based on bitter cassava starch (*Manihot esculenta*) and chitosan extracted from crab shell (*Scylla seratta*). This consists in analyzing the influence of chitosan concentration on the physicochemical, mechanical and water barrier properties of the starch film.

## II.MATERIALS AND METHODS

### 1. Materials

Starch with a moisture content of 9,89% of total weight extracted from bitter cassava and chitosan with purity degree of 99,99% derived from crab shells were used to develop the biodegradable films. Glacial acetic acid, distilled water, and glycerol as plasticizer.

### 2. Methods

#### 2.1. Preparation of starch from bitter cassava

Cassava tubers without visible defects are washed, peeled, rinsed with distilled water and grated. Distilled water is then added to the whole, which is crushed and macerated for 10 to 15 minutes. The liquid phase is separated from the tuber pulps by filtration. This operation is repeated 5 times until the liquid flowing out of the filtration is clear. The collected liquid called "starch milk" is then decanted for 6 hours. The supernatant is removed and the deposited sediment is dried at 30°C in oven for 72 hours. The lumps obtained are then crushed and sieved. The powder starch is obtained and ready for use as a bioplastic feedstock (Figure 1).

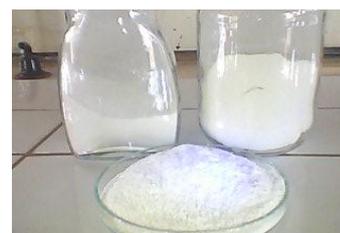

Figure 1: Cassava starch powder

#### 2.2. Production of chitosan from crab shells
#### 2.2.1. Preparation of crab shells before extraction

Fresh crabs were collected from local market of Antananarivo. Crabs are shelled and their shells are then washed with tap water and then with distilled water. In order to remove all traces of impurities such as flesh remains, the shells have been boiled over low heat for a few minutes and then dried in an oven at a temperature of 90°C for 24 hours (Figure 2).

After drying, shells are cooled then crushed with a grinder and sieved to obtain fine shell powder.

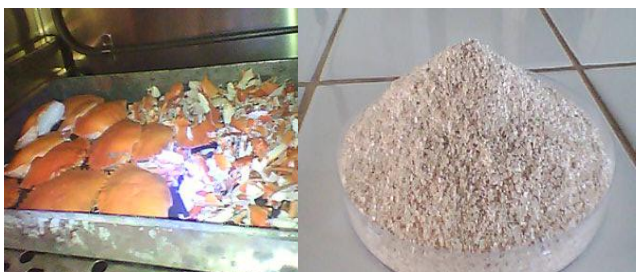

Figure 2 Dried crab shells and fine shell powder

### 2.2.2.  Isolation of chitosan

The following 4 steps namely: Demineralization, Deproteinization, Discoloration or Bleaching, and Deacetylation are followed for the production of chitosan from crab shells [8].

#### Demineralization:

Demineralization was carried out by adding 450 mL of 2M HCl to 90 g of crab shell powders under agitation at temperature 85°C for 3 hours. Afterwards, the demineralized crab shells were filtrated and washed in distilled water until neutral pH, to remove all acid traces.

#### Deproteinization:

Deproteinization of demineralized crab shells was performed with 2M NaOH at a solid to solvent ratio of 1:10 (g/mL). Reaction was carried under agitation at 100°C for 3 hours. The solid was filtrated and soaked in distilled water until neutral pH, to remove alkaline traces and then filtrated. This is how chitin is obtained (Figure 3).

#### Discoloration or bleaching:

The resulting chitin after deproteinization is bleached with 0,32% NaOCl (sodium hypochlorite solution) at a solid to solvent ratio of 1:10 (g/mL) for 3 minutes at ambient temperature. The discolored chitin is then filtrated, washed with distilled water and dried in an oven at 60°C for 4 hours.

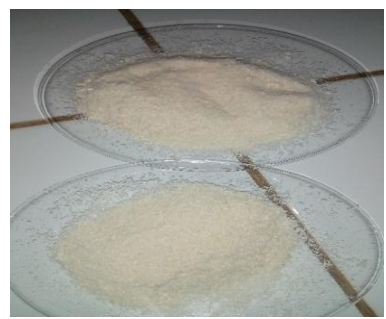

Figure 3: Extracted chitin

#### Deactivation

The transformation of chitin into chitosan consists essentially in eliminating the acetyl groups linked to the amine group of the different chitin units. Deacetylation of chitin was experimented by reacting chitin with 12,5M NaOH at a solid to solvent ratio of 1:10 (g/mL) at 100°C under agitation for 3 hours. The residue was filtrated and washed with distilled water until neutral pH to remove alkaline traces. The resulting chitosan was dried at 60°C in an oven for 4 hours. Then the chitosan is ready to use for the elaboration of bioplastic films (Figure 4).

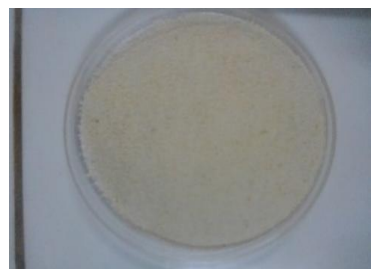

Figure 4 Extracted chitosan

### 2.3.  Bioplastic synthesis

Casting method was employed to prepare bioplastic films.

### 2.3.1.  Preparation of thermoplastic starch (TPS)

Cassava starch solution was prepared by mixing 10g of cassava starch and 100 mL of distilled water. An amount of glycerol is added with a volume of 0,2mL/g according to the total mass of starch in solution. And heating the mixture on hot plates with stirring until they gelatinized at 80°C for 15 minutes. The mixture is poured up to 25 mL into a Petri dish (diameter: 8,5 cm) and dried in an oven at 35°C for 48h. This TPS without addition of chitosan was also prepared and used as a control.

### 2.3.2. Preparation of thermoplastic Chitosan (TPC)

Chitosan-based films were obtained by dispersing 2g of chitosan in an aqueous solution of glacial acetic acid with concentration of 1%. 25 mL of the solution was then poured consistently into a Petri dish (diameter: 8,5 cm) and dried in an oven at 35°C for 48h.

### 2.3.3. Preparation of starch/chitosanbioplastics

Bioplastic films was performed by weighing cassava starch and chitosan with varying predetermined mass.

The different rates of the starch/chitosan mixture, in percentage by weight according to various ratios are: 50/50; 70/30; 80/20; 90/10 (weight/weight) at a rate of 10 grams of total weight of starch/chitosan, as illustrated in Table 1.

Table1: Different rates of starch/chitosan mixture

| Ratio of Cassava Starch to chitosan (%) | Cassava starch mass (g) | Chitosan mass (g) | Total mass of biopolymers (g) |
|---|---|---|---|
| Control (TPS) | 10 | 0 | 10 |
| S90/C10 | 9 | 1 | 10 |
| S80/C20 | 8 | 2 | 10 |
| S70/C30 | 7 | 3 | 10 |
| S50/C50 | 5 | 5 | 10 |

Starch solutions at concentrations of 5, 7, 8, and 9% (g/mL) were prepared by dispersing cassava starch, in distilled water and heating the mixtures on hot plates with stirring at 80°C until they gelatinized and then cooling to 25°C. Chitosan solution at different concentration of 1, 2, 3, 5 % (g/mL) (10, 20 ,30, and 50% of dry starch solid weight) were dispersed in an aqueous solution of glacial acetic acid at concentration of 1mL/100mL under stirring until the chitosan dissolved completely. A series of starch/chitosan composite films were prepared by mixing 100 mL of 5% starch solution with 100mL of 5% chitosan solution; 100 mL of 7% starch solution with 100mL of 3% chitosan solution; 100 mL of 8% starch solution with 100mL of 2% chitosan solution; 100 mL of 9% starch solution with 100mL of 1% chitosan solution.

Then the mixtures are put in a water bath at 80°C under stirring for 25 minutes and glycerol is added with a volume of 0,2mL/g according to the total mass of polymers in solution, under stirring until homogeneity. The compositions of mixtures are given in Table 2. When the mixtures have become homogeneous, they are taken out of the water bath, then the hot film-forming solutions are poured up to 25 mL into petri dishes (diameter: 8,5 cm) and dried in the oven with air circulation for 48 hours at 35°C. Dried bioplastic films were peeled off manually using spatula and store in the desiccator prior to characterization.

Table2 Compositions of starch/chitosan films

| | | Compositions | | |
|---|---|---|---|---|
| Starch solution (g/mL) | Chitosan solution (g/mL) | Volume of starch solution (mL) | Volume of chitosan solution (mL) | Volume of the glycerol (mL) |
| 9% | 1% | 100 | 100 | 2 |
| 8% | 2% | 100 | 100 | 2 |
| 7% | 3% | 100 | 100 | 2 |
| 5% | 5% | 100 | 100 | 2 |

## 3. Characterization of bioplastic films
### 3.1. Film thickness

Thickness of the bioplastic films was measured with a precision digital micrometer (Teclock, model SM-112/ Japan) with sensitivity of 0,01mm. Five thickness measurements were taken on each film, one in the center and four near the perimeter of the film. The average value of the thickness was used to calculate the water vapor permeability (WVP), tensile strength (TS), elongation at break (E), and Young's Modulus (YM).

### 3.2. Moisture content (MC)

The moisture content (%) of bioplastic films was determined as per the gravimetric method.

The samples were dried in an oven at 105°C for 6 hours until the equilibrium weight was attained. Water content was calculated as the percentage of initial film weight loss through drying according to the equation (1) [9].

$$MC = \frac{W_i - W_f}{W_i} \times 100 \qquad (1)$$

$W_i$: initial weight of film
$W_f$: final weight of film

### 3.3. Water solubility (WS)

Film pieces (diameter: 6cm) were dried at 105°C in an oven for 6 hours for the initial dry mass. Films were immersed into 40mL of distilled water for 24 hours of immersion at 25°C with agitation. After that, pieces of bioplastic film were filtered and dried till constant weight achieved in an oven at 105°C.
The water solubility (%) of films was calculated from the following equation (2)[9].

$$WS = \frac{m_i - m_f}{m_i} \times 100 \qquad (2)$$

$m_i$: mass of initial dry film
$m_f$: mass of undissolved film

### 3.4. Water absorption (WA)

Dried bioplastic films were weighted with an analytical balance sheet and soaked in distilled water at room temperature for 2 hours. Every 30 minutes, the samples were taken and the water on the films surface were wiped with a cloth and weight again. Water absorption was calculated using equation (3)[10][11].

$$WA = \frac{m_f - m_i}{m_i} \times 100 \qquad (3)$$

$m_f$: mass final of film after immersion in distilled water
$m_i$: mass initial of dry film

### 3.5. Water vapor permeability

Determination of water vapor permeability of film samples was determined using desiccant method according to ASTM E96/E96Mmethod[12], with some modifications. Circular film samples were placed hermetically on the open mouth of cylindrical cup, with 4,5cm of internal diameter (exposed area: 15,90 $cm^2$) containing 30 ml of distilled water with a relative humidity (RH) fixed at 100%. The assembly was weighed and placed into a desiccator containing silica gels with 0% relative humidity at 25°C. Each sample was periodically taken out and weighed until a constant weight was obtained. Five weight measurements were made over 4h for 20h. Changes in the weight of the cups were recorded using an analytical balance with a precision of 0,0001g and plotted as a function of the time.

The slope of each line represented the amount of water vapor passing through the film per unit of time. Which was calculated by linear regression. The water vapor transmission rate WVTR ($g.m^{-2}.s^{-1}$) was determined from de slope $\frac{\Delta m}{\Delta t}$ ($g.s^{-1}$) of the straight line divided by the exposed area ($m^2$) and calculated according to the equation (4)[13].

The water vapor permeability of films was determined using the equation (5)[14][15].

$$WVTR = \frac{\Delta m}{\Delta t \times A} \qquad (4)$$

$$WVP = \frac{WVTR \times T}{P_0(RH_1 - RH_2)} \qquad (5)$$

Where $\Delta m$ is the change in mass of the permeation cell for a time $\Delta t$, A is the exposed area, T is the average thickness of the film, $P_0$ is the saturation vapor pressure of water at 25°C, RH1 is the relative humidity n the permeation cell, RH2 is the relative humidity in the desiccator.

### 3.6. Fourier transform Infrared Spectroscopy (FTIR)

FTIR absorption analysis was carried out by using an IR spectrometer (Shimadzu model 8400 S) to determine the FTIR spectra at the frequency range of $4000 - 400$ $cm^{-1}$.FTIR analyses were performed to study the effect of the addition of chitosan in thermoplastic starch and to verify possible interaction among cassava-starch, chitosan. The samples were prepared using KBr-disk method[16].

### 3.7. Mechanical properties

Tensile strength TS, maximum elongation at break E and Young's Modulus were measured with a universal testing machine (Kao Thiech, Model KT-7010-C, Taichung Taiwan), following the ASTM standard method[17], with some modification. The specimens were 30mm wide, 70mm long and 0,218mm in thickness, which were mounted in the film extension grips.

Three samples were tested for each formulation. Specimens were strained at a constant rate of 50mm.min$^{-1}$ until rupture.

The tensile strength was calculated by equation (6) [18].

$$TS = \frac{F}{S} \qquad (6)$$

Where, F is the maximum ultimate breaking force (N), S is the cross-sectional area (m$^2$).

The elongation at break was determined by equation (7) [18].

$$E = \frac{l_f - l_0}{l_0} \qquad (7)$$

Where, $l_f$ is the final sample length and $l_0$ is initial sample length.

While Young's modulus determined using the following formula:

$$YM = \frac{TS}{E} \qquad (8)$$

Where TS is the tensile strength and E is the elongation at break.

### 3.8. Biodegradation test

Biodegradability of the bioplastic films was detected by placing the samples in a natural outdoor environment for four weeks in order to observe and follow their degradation processes.

This test aims to verify if the synthesized bioplastics were biodegradable without being buried in the soil. therefore, specimens were placed on the soil surface and weighed weekly to evaluate the degradation rate.

Weight loss percentage of the samples was determined by employing the equation (9)[19].

$$Weight\,loss = \frac{w_i - w_f}{w_i} \qquad (9)$$

$w_i$ is the weight of sample before and $w_f$ is the weight of the sample after degradation.

## III. RESULTS AND DISCUSSIONS

### 1. Bioplastic films appearance

Films obtained in this study were homogeneous, flexibles and did not exhibit any cracks, while the thermoplastic starch (TPS) was transparent, the thermoplastic chitosan (TPC) and the starch/chitosan blend films were slightly colored plastic films, as shown in Figure 5. The resulting of starch/chitosan blend films color was influenced by the chitosan color.

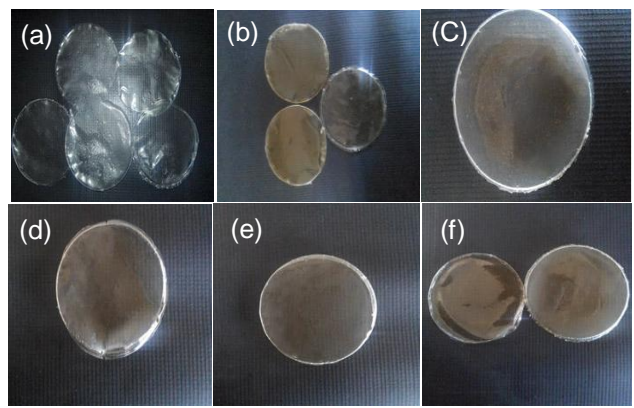

Figure 5: Cassava starch films, chitosan films, and cassava starch-chitosan blend films. (a) Control (TPS: pure cassava starch films), (b) chitosan films (TPC), (c) cassava starch/chitosan (90/10), (d) cassava starch/chitosan (80/20), (e) cassava starch/chitosan (70/30), (f) cassava starch/chitosan(50/50).

### 2. Films thickness

The thickness of bioplastic films produced was set between 216 μm to 218 μm as shown in the Table 3. There was no significant increase in the neat cassava starch-based films (216 μm) and the neat chitosan films (217 μm) thickness when compared with the starch-chitosan blend films.

Values of films thickness was used to calculate their water vapor permeability and mechanical properties. It has been proven that thickness affects properties of the film as well as the barrier and mechanical properties [20].

Table 3 Physical properties of cassava starch/chitosan composite films with different concentration with chitosan

| Samples | Thickness (μm) | Moisture content (%) | Water solubility (%) | Water absorption (%) | Water vapor permeability (g.s$^{-1}$.m$^{-1}$.Pa$^{-1}$) |
|---|---|---|---|---|---|
| Control (TPS) | 216 | 8,3911 | 11,8329 | 91,3545 | 9,5540.10$^{-12}$ |
| TPC | 217 | 7,4257 | 11,3225 | 24,4815 | 8,3984.10$^{-12}$ |
| S90/C10 | 218 | 6,0433 | 11,2115 | 56,5454 | 7,2318.10$^{-12}$ |
| S80/C20 | 218 | 5,8618 | 11,1031 | 53,4568 | 6,0265.10$^{-12}$ |
| S70/C30 | 217 | 5,4453 | 10,2113 | 51,0391 | 4,7991.10$^{-12}$ |
| S50/C50 | 218 | 5,0215 | 10,0751 | 50,4568 | 2,4106.10$^{-12}$ |

## 3. Moisture content

The knowledge of moisture content of the bioplastic films is very important for the evaluation of the solubility and the water vapor transmission of the films. The higher moisture content in films makes them more hydrophilic and results in low water permeability [9] [21]. Thus, the amount of water present in bioplastic films acting as a solvent favors chemical or biochemical reactions and also leads to the growth of microorganism [9]. As can be seen in Table 3, cassava-starch films (control) and pure chitosan films show higher moisture content values than bioplastic films obtained from chitosan containing films.

The decrease of moisture content values when cassava starch was incorporated with chitosan can be correlated with a good interaction between starch and chitosan. Such attributable to the formation of strong bonds between hydroxyl groups of starch and the amino groups of chitosan, indicated that the number of active sites for water binding decreased with increasing chitosan ratio in the film components. However, the same tendency was observed by Y. Zhao et al. [9].

## 4. Water solubility

The solubility of films in water is an important property of edible films that give additional indication for film's water affinity and information to select the films for specific applications.
The results for water solubility of the films ranged from 10,0751% to 11,8329% (control) as tabulated inTable 3. It was found that, incorporation of chitosan reduced water solubility of starch-based films (Figure 6).

Prior studies reports that chitosan has higher hydrophobicity when compared to starch[22]. Hydrophobicity property of chitosan could be responsible for the existence of fewer interaction between films matrix and water [22].

Additionally, formation of the intermolecular hydrogen bonds interactions between cassava starch and chitosan molecules reduced thenumber of the hydrophilic groups that could be interact with water molecules. Therefore, chitosan addition to starch reduced film solubility in water.Hydrophilic content decreased as a result of incorporation of chitosan.

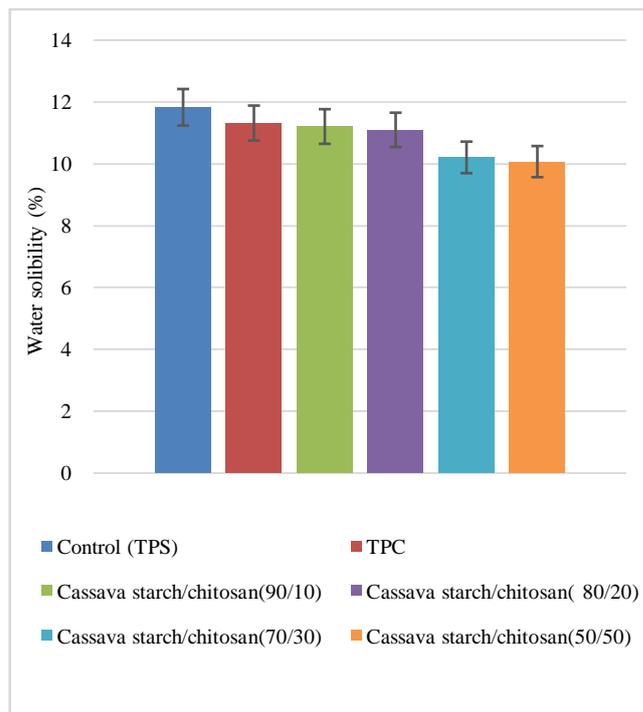

Figure 6: Water solubility of pure starch film, pure chitosan film and cassava starch/chitosan films.

## 5.Water absorption

Water absorption test is carried out to evaluatehydrophilicity, water absorption capacity of the film surface and to determine the resistance of bioplastic samples in water. The effect of chitosan content on water absorption of plasticized starch-based films and the percentage of water uptake as a function of time in cassava starch/chitosan composite films are shown in Figure7. Based on the data in Table 3, water absorption values decreased as the content of chitosan increased. The highest absorption capacity of water is produced by cassava starch-based films (TPS).

Thermoplastic starch was more sensitive to water than thermoplastic chitosan because of its hydrophilic nature than chitosan. However, cassava starch/chitosan composite films show hydrophobic properties compared to the starch-based films. The lowest absorption capacity of water was obtained by cassava starch/chitosan blend films at a 50/50 composition.The addition of chitosan has the capability to decrease water absorption. This is due to the basic nature of chitosan that its higher hydrophobicity, so that in a balanced proportion between cassava starch polymer and chitosan polymer cause hydrogen bond interactions that can decrease the hydrophilic nature of cassava starch [23] [24].

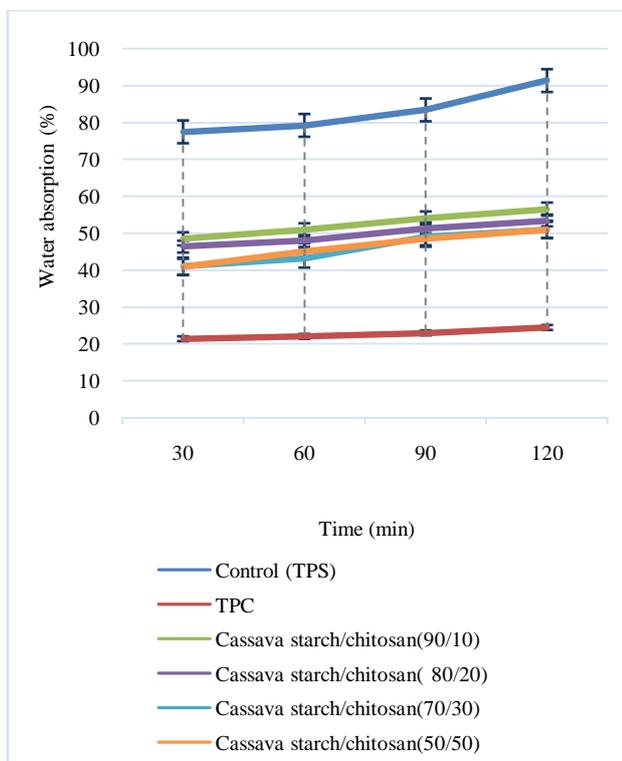

Figure 7: Water absorption of TPS and cassava starch/chitosan composite films with different concentrations of chitosan.

## 6.Water vapor permeability

The values of water vapor permeabilityof the composite films as a function of chitosan ratio are shown in Table 3. Following Figure 8 shows that the water vapor permeability of cassava starch-based film was $9,5540.10^{-12} g.s^{-1}.m^{-1}.Pa^{-1}$, while those of cassava starch/chitosan composite films were in the range of $2,4106.10^{-12}$ to $7,2318.10^{-12}$ $g.s^{-1}.m^{-1}.Pa^{-1}$.

It was shown that the water vapor permeability (WVP) of the films with the addition of chitosan was significantly lower than without the additive. This result indicated that adding chitosan could improve the water vapor properties of cassava starch-based film. This behavior is probably due to the intermolecular hydrogen bond formation between cassava starch and chitosan molecules and resulting in the reduced of number of free hydroxyl groups that could interact with water[25] [26].

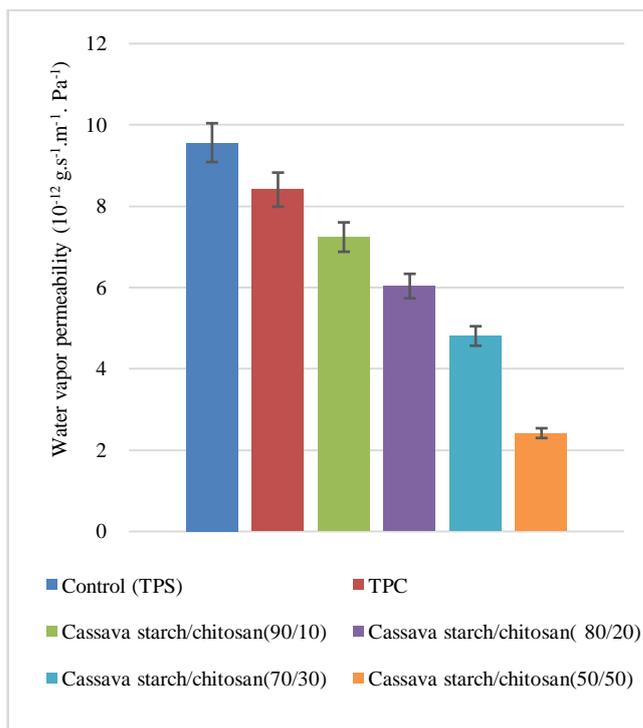

Figure 8: Water vapor permeability of TPS and cassava starch/chitosan composite films with different concentrations of chitosan.

## 7.Mechanical properties

The mechanical properties (tensile strength, elongation to break, Young's modulus) of bioplastic samples are summarized in Table 4.

Table4: Mechanical properties (tensile strength (TS), elongation to break (E), Young's modulus (YM) of the cassava starch film, chitosan film, cassava starch/chitosan films.

| Samples | TS (MPa) | E (%) | YM (MPa) |
|---|---|---|---|
| Control (TPS) | 2,2708 | 50,0000 | 4,5416 |
| TPC | 0,6028 | 21,4285 | 2,8131 |
| S90/C10 | 3,4500 | 52,8571 | 6,5270 |
| S80/C20 | 4,5000 | 54,2857 | 8,2895 |
| S70/C30 | 5,2742 | 64,2857 | 8,2043 |
| S50/C50 | 6,3000 | 62,8571 | 10,0227 |

From the Table 4 and Figure 9, it can be observed that the tensile strength (TS) of the cassava starch film increases from 2,2708 to 6,3000 MPa with increasing chitosan concentrations. The lowest tensile strength (2,2708 MPa) was recorded for thermoplastic starch (control), whereas the highestresistance (6,3000 MPa) was recorded for bioplastic in the composition of starch/chitosan 50/50. According tothe results, the tensile strength of the biofilms was significantly affected by the addition of chitosan.

There was an improvement in tensile strength compared to the neat cassava starch film without the addition of chitosan.The increasing tensile strength values of the cassava starch/chitosan films might be a result of the formation of intermolecular hydrogen bonds between OH$^-$ of the cassava starch and NH$_3^+$ of the chitosan, which are supported by the results of FTIR. The amino groups (NH$_2$) of the chitosan were protonated to NH$_3^+$ in the glacial acetic acid solvent, whereas the ordered crystalline structures of the cassava starch molecules were destroyed during gelatinization process, resulting in OH$^-$ groups being exposed to readily from hydrogen bond with NH$_3^+$ of the chitosan polymer, facilitating the reaction between chitosan and cassava starch [9] [27].

Some previous studies also revealed that increased tensile strength values with increasing chitosan content was attributed to the formation of intermolecular hydrogen bonding between the hydroxyl (-OH) groups of the cassava starch and the amino (-NH$_2$) groups of chitosan [4] [28].
Elongation at break (E) is an important parameter to determine the film's flexibility and elongation capacity before the film could break [29]. Table 4 and Figure 10 show that the addition of chitosan significantly affected the elongation at break.It was found thatthe elongation at break of cassava starch/chitosan films increased with incorporation of chitosan at concentration between 10 to 30%,as compared to neat cassava starch film.

The maximum (64,2857%) occurred at the cassava starch/chitosan ratio of 70/30. However, when the chitosan ratio was increased further to 50%, the elongation at break values suddenly decreased to62,8571%. The addition of too much chitosan lowered the flexibility of the film.The increased flexibility of the films at increased chitosan concentration could have been provoked by the interaction of plasticizer-polymer chains while improving the chain mobility, the elongation at break as well as the flexibility of films [30].Additionally, the decreased flexibility of the films with higher chitosan concentration might be linked to the number of NH$_3^+$ groups went up with increased chitosan proportion in the film forming solution and when the concentration exceeded a critical value, it is very difficult to form homogeneous starch-chitosan mixture[14], resulting a decreased elongation at break of the blend films. A similar behavior was reported in some previous studies[14][31],in which theinfluence of chitosan concentration on mechanical and barrier properties of corn starch/chitosan films, and biodegradable polymer blends based on corn starch and thermoplastic chitosan processed by extrusion were studied.

Young's modulus is a useful parameter that determines the film stiffness. The lower Young's modulus value indicates that the more flexible bioplastic samples, the higher Young's modulus value indicates the film has less elasticity [14]. Figure 11shows the effect of chitosan addition on the Young's modulus of bioplastic films. It is shown that the addition of 10% to 50% of chitosan significantly increased the Young's modulus of the film compared to Young's modulus value of the neat cassava-starch based film. The rise in Young's modulus value was due to the increased tensile strength value.

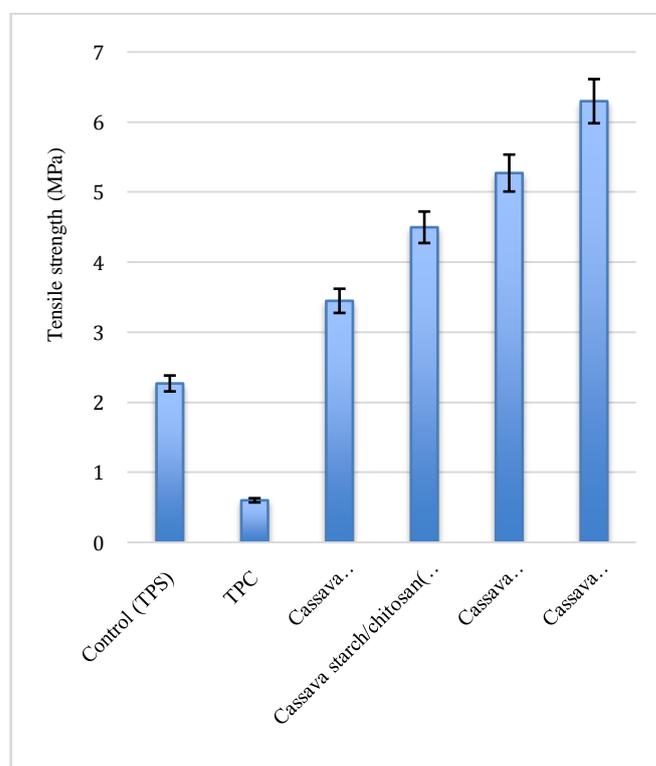

Figure 9: Tensile strength of cassava starch films with different cassava starch/chitosan ratios.

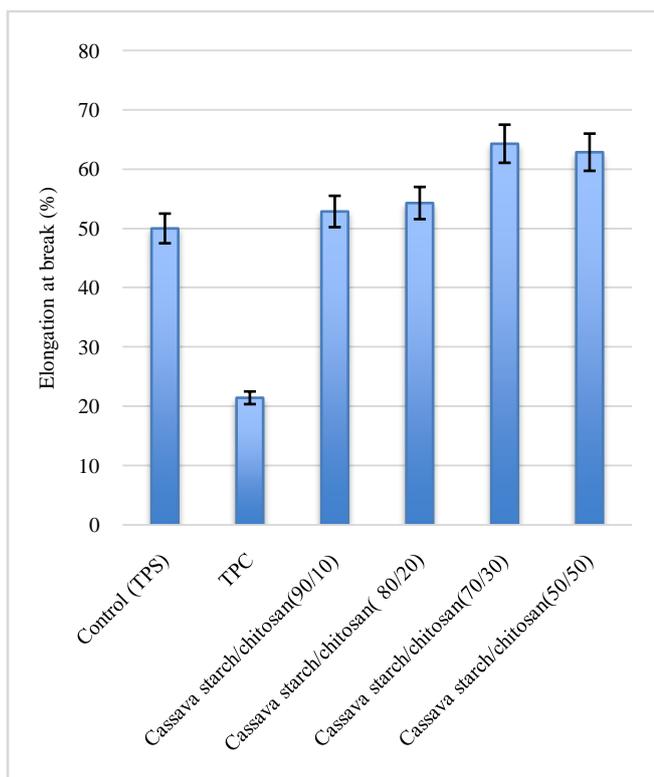

Figure 10: Elongation at break of cassava starch films with different cassava starch/chitosan ratios.

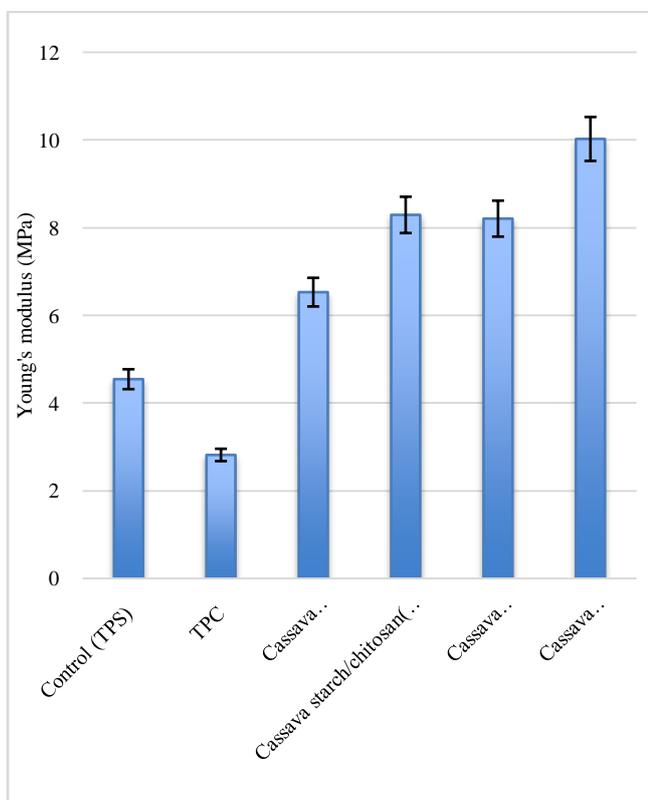

Figure 11: Young's modulus of cassava starch films with different cassava starch/chitosan ratios.

## 8. Fourier Transform Infrared spectroscopy (FTIR)

FTIR spectroscopy was applied to verify and to examine the chemical interaction between cassava starch and chitosan. The FTIR spectra of cassava starch film, chitosan film, and cassava starch-chitosan blend film are shown inFigure 12, Figure 13, Figure 14, respectively. As shown in Figure 12,13,14, the typical region of saccharide bands between 1180 cm$^{-1}$and 953 cm$^{-1}$ which is often considered to comprise vibration modes of C-O and C-C stretching and the bending mode of C-H bonds [32].

In the spectra of cassava starch film (Figure 12), typical absorption band at 3321 cm$^{-1}$ is attributed to the hydrogen bonded hydroxyl groups (O-H stretching)associated with free inter and intramolecular bound hydroxyl group[33]. The sharp band at 2872 cm$^{-1}$ is characteristic of C-H stretching vibration. The third absorption band at 1630 cm$^{-1}$ is attributed to water bending $\delta$(O-H) due to the presence of bond water in the film [32] [33]. Finally, the characteristic absorption bands at 1160 cm$^{-1}$ and 1020 cm$^{-1}$are attributed from the C-O in C-O-H and C-O in C-O-C chemical bonds, respectively [34].

In the spectra of the chitosan film (Figure 13), the absorption band at 3268 cm$^{-1}$ corresponding to the stretching vibration of N-H and hydrogen-bonded hydroxyl groups [14] [15]. The peak located at 2832 cm$^{-1}$ was attributed to CH$_2$- stretching vibration on pyranose ring[15]. The absorption peak appearing at 1632 cm$^{-1}$ was associated with the stretching of C=O (amide-I), and the peak located at 1550 cm$^{-1}$ corresponding to the absorption band of N-H bending(amide II) [14].

When cassava starch and chitosan are mixed, physical blends versus chemical interactions are reflected by changes in the characteristic spectra peaks[27]. As can be seen in Figure14, compared to the IR spectra of pure cassava starch film (TPS) and chitosan film (TPC), the characteristic peak of inter and intra-molecular hydrogen bonds in cassava starch film at 3321 cm$^{-1}$ and in chitosan at 3268 cm$^{-1}$ shifted to a higher wavenumber at 3329 cm$^{-1}$ in cassava starch-chitosan blend film. Additionally, the amide I and N-H bending characteristic peak of chitosan in the IR spectra of cassava starch-chitosan composite film shifted from 1632 cm$^{-1}$ to 1641 cm$^{-1}$ and 1550 cm$^{-1}$ to 1555 cm$^{-1}$. Similar shifts were also observed in other previous studies [14][27] [32].

These results of FTIR analysis indicated that there was formation of inter and intra-molecular hydrogen bonding between hydroxyl groups of cassava starch and the amino groups of chitosan[27] [35].

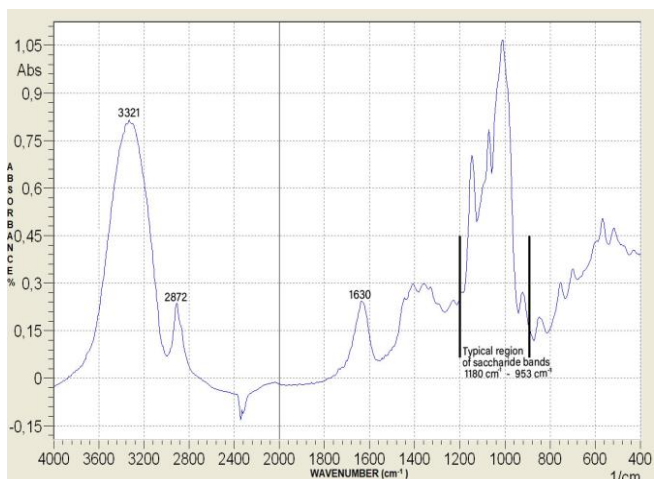

Figure 12: FTIR spectra for cassava starch film (TPS).

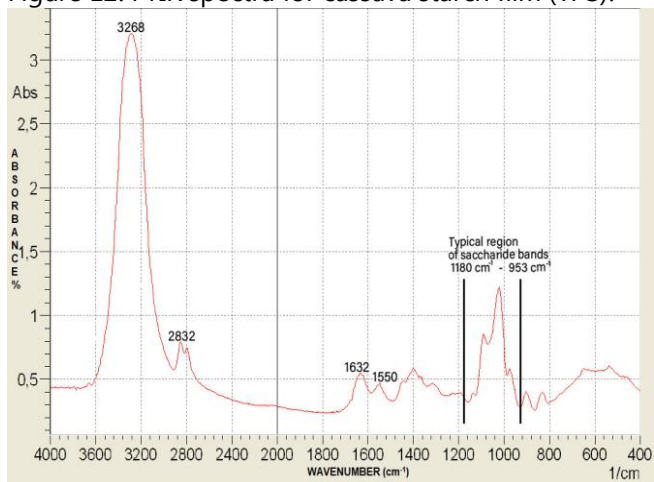

Figure 13: FTIR spectra for chitosan film (TPC).

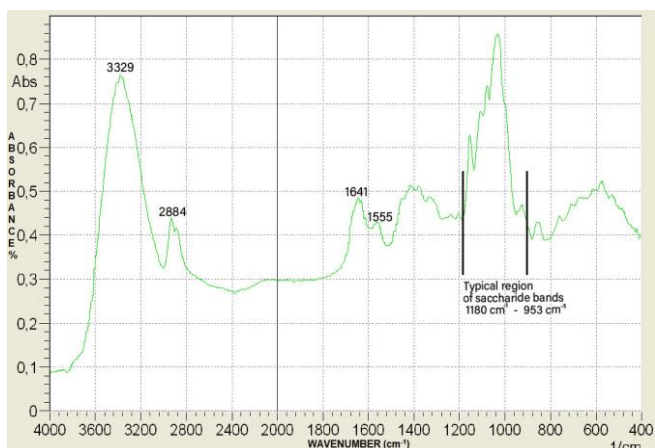

Figure 14: FTIR spectra for cassava starch-chitosan blend film.

## 9. Biodegradation test

An important advantage of films prepared with biodegradable materials is related to the reduction of environmental pollution. Biodegradation test results of bioplastic films carried out for 1, 2, 3 and 4 weeks can be seen in Figure15. It is shown that biodegradation rate increased with the increasing degradation and increasing time. From the Figure 15, it is observed that bioplastic films based on cassava starch and chitosan mixture degraded slower than the pure cassava starch-based film after four weeks of exposure to the soil surface.That shows that the addition of chitosan was slowed the biodegradation of cassava starch-based films. This might be attributed to the antibacterial property of chitosan because chitosan has antibacterial activity against Gram-negative bacteria and Gram-positive bacteria [36]and has been used for production of antimicrobial packaging films [37] [38].

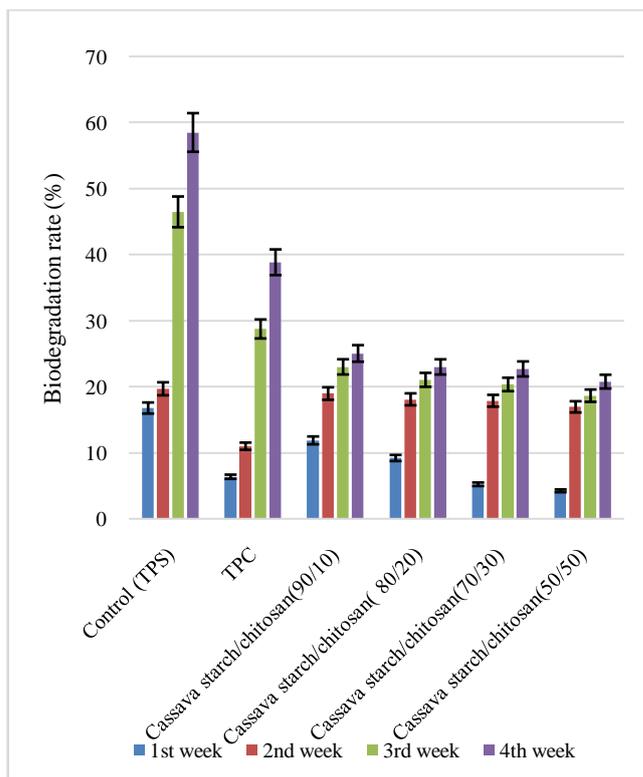

Figure 15: Biodegradation test for Cassava starch films, chitosan films, and cassava starch-chitosan blend films.

According to this biodegradation test results, it is expected that these films are biodegradable and compostable on the soil surface in less than six months.In cooperation with moisture, microorganism decomposers and solar radiation, the structure of film fibers decomposes, which ultimately results in biodegradation. Additionally,in the prepared films, cassava starchand chitosan, are all natural polymers and totally biodegradable.Therefore, the films did not lose its total biodegradable character after mixing but withstand its stability for longer period.

## IV.CONCLUSION

The effects of chitosan concentration (0%,10%, 20%,30%,50%) on physicochemical, mechanical and water barrier properties of bioplastic films based on cassava starch were comprehensively studied. It can be concluded by FTIR results that chitosan could form hydrogen bond interaction with cassava starch. The incorporation of chitosan improved strength, stiffness, and reduced water permeabilityof the cassava starch-based film. This study demonstrates an effective strategy to improve the performance of cassava starch film.The development of cassava starch-chitosan blend film for use as active packaging films turns out to be promising. Theyhave potential applications by their biodegradable characteristic that offer tangible advantages.